# High-dimensional Bid Learning for Energy Storage Bidding in Energy Markets

Jinyu Liu[1], Hongye Guo[1], Qinghu Tang[1], En Lu[2], Qiuna Cai[2], Qixin Chen[1*]

1 Department of Electrical Engineering, Tsinghua university, Beijing, 100084, China

2 Guangdong Power Grid Corporation Power Dispatching & Control Center, Guangzhou, 510335, China

**ABSTRACT**

With the growing penetration of renewable energy resource, electricity market prices have exhibited greater volatility. Therefore, it is important for Energy Storage Systems(ESSs) to leverage the multidimensional nature of energy market bids to maximize profitability. However, current learning methods cannot fully utilize the high-dimensional price-quantity bids in the energy markets. To address this challenge, we modify the common reinforcement learning(RL) process by proposing a new bid representation method called Neural Network Embedded Bids (NNEBs). NNEBs refer to market bids that are represented by monotonic neural networks with discrete outputs. To achieve effective learning of NNEBs, we first learn a neural network as a strategic mapping from the market price to ESS power output with RL. Then, we re-train the network with two training modifications to make the network output monotonic and discrete. Finally, the neural network is equivalently converted into a high-dimensional bid for bidding. We conducted experiments over real-world market datasets. Our studies show that the proposed method achieves 18% higher profit than the baseline and up to 78% profit of the optimal market bidder.

**NONMENCLATURE**

| | |
|---|---|
| *Abbreviations* | |
| ESS | Energy Storage System |
| NNEB | Neural Network Embedded Bid |
| RL | Reinforcement Learning |
| *Symbols* | |
| $\lambda$ | Market Price |
| $p$ | ESS Power Output |

## 1. INTRODUCTION

In recent years, the installed capacity of Energy Storage Systems (ESSs) has witnessed a substantial growth. Consequently, the strategic participation of ESS in electricity markets has become a hotspot of research[1,2]. At the same time, with the growing generation of renewable energies, market prices have become more and more unpredictable and volatile, so it becomes crucial to utilize effective energy bids to respond wisely to market prices.

Driven by the developments in machine learning communities, RL based methods have been widely adopted in the strategic bidding[1] and self-scheduling of ESSs. However, current methods hardly fully utilize the high-dimensional energy bids to hedge market uncertainties. In most spot markets, a Generation Company(GenCo) submits 10-10 monotonic price-quantity pairs for market clearing[3]. This high-dimensional bidding space is flexible for hedging uncertainties, but its high dimensionality makes it unsuitable for the current RL methods used in this area.

Current RL-based methods use two kinds of simplifications to tackle the high-dimensional(10-10) bid format challenge. The first kind of simplification directly changes the market rule to the format of low-dimensional bids, which could be one price-quantity pair[1], or only one price bid with fixed quantity[4]. The second type of simplification reparametrize the high-dimensional bids to low-dimensional spaces like 1-dim overflow proportion[5], 1-dim supply slope[6] or other custom spaces[7]. Both types of simplification cause the bid expressiveness to be limited, which would further hinder the profits from bidding. Thus, it is important to develop an effective method to better utilize the high-dimensional bid space.

In order to tackle this challenge, we propose the NNEB as a bid representation method to generate high-dimensional bids by learning a neural network. NNEB refers to the market bid that is represented by a neural network which have monotonic and discrete outputs. First, we demonstrate that bids can be considered as a special type of supply function which is monotonic and discrete. Then, we utilize RL algorithms to learn a neural network as a strategic supply function. Next, we satisfy network monotonicity by proposing a new training loss term and satisfy discreteness with output activation. Finally, the monotonic and discrete neural network is transformed back into high-dimensional(10-10) bids for storage arbitrage.

The rest of this paper is structured as follows: In Section 2, we propose the bidding problem by defining the ESS model and market model. In Section 3, we propose our methodology for constructing NNEBs. In Section 4, we conduct experiments on ESS units using PJM market histories. Finally, the concluding remarks are provided in Section 5.

## 2. PROBLEM FORMULATION

### 2.1 ESS Operation Model

To model the system transitions, we first provide the energy storage operation model, it mainly consists of two parts: the storage capability model and the life depreciation model.

The storage capability model tracks the evolution of the State of Charge (SoC). It can be described in time-step formats:

$$SoC_t = SoC_{t-1} + \tau(\eta^c p_t^c - \frac{p_t^d}{\eta^d})$$
$$0 \leqslant SoC_t \leqslant SoC_{\max},$$
$$0 \leqslant p_t^c \leqslant p_{\max}, \quad (1)$$
$$0 \leqslant p_t^d \leqslant p_{\max},$$
$$p_t^c \cdot p_t^d = 0$$

where $\tau$ is the incremental timestep, usually 5 minutes for real-time markets. $\eta^c, \eta^d$ represents the charging and discharging efficiency, and $p_t^c, p_t^d$ is the charging and discharging power.

The lifespan model of the ESS describes its operational costs. We utilize a depreciation model whereby the ESS's degradation cost is proportional to the power output and time elapsed. The model can be expressed formally as:

$$r^{dep} = -\tau \cdot \lambda^{dep} \cdot |p_t| \quad (2)$$

### 2.2 Market Model

We propose the real-time market based on the bidding process in PJM[3]. The real-time market operates through an hourly bidding process and a 5-minute clearing process. At the beginning of each operational hour, market participants submit energy bids. The market price is then determined by clearing the bids for the following twelve 5-minute periods.

In the energy market, bids are typically presented as a series of 10 price-quantity pairs that monotonically increase. A valid bid consists of the price bids $\lambda_{t,i}, i = 1,\ldots,10$, and the quantity bids $p_{t,i}, i = 1,\ldots,10$.

As an ESS has a small capacity, it is regarded as a price-taker in market clearing. Therefore, its cleared power is determined by the clearing price:

$$p_t(\lambda_t) = \begin{cases} p_{\max} & \text{if } \lambda_{t,10} \leqslant \lambda_t \\ p_{t,i} & \text{if } \lambda_{t,i} \leqslant \lambda_t < \lambda_{t+1,i} \\ p_{\min} & \text{if } \lambda_t < \lambda_{t,1} \end{cases} \quad (3)$$

The final net income of the ESS, is the market payment subtracted by the storage depreciation cost:

$$r_t^{net} = \tau(\lambda_t \cdot p_t - \lambda^{dep} \cdot |p_t|) \quad (4)$$

which is the *net income reward* used in RL training.

The aim of our proposed methods is to maximize the cumulative net income (4) with proper formulations.

## 3. METHODOLOGY

### 3.1 Neural Network Embedded Bids

NNEBs refer to high-dimensional bids that are represented by monotonic and discrete neural networks. As shown in *Fig. 1*, a high-dimensional bid can be considered as a special type of market supply function which are monotonic and discrete. Therefore, they can be approximated by neural networks if the network input-output relation is both monotonic and discrete. When both of these constraints are met, a valid bid and a neural network $p_t = \pi_{\theta,o_t}(\lambda_t)$ are in one-to-one correspondence.

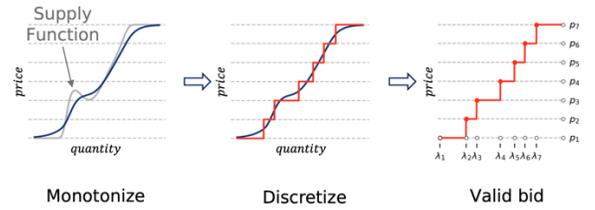

Fig. 1 Factorization of bid constraints

To learn and generate NNEBs, we will first use RL to learn a neural network as the bidder's supply function in the next section, then satisfy the two constraints by proposing learning modifications in Section 3.3 and 3.4.

### 3.2 Supply Function Learning

We utilize the Proximal Policy Optimization(PPO)[8] RL algorithm to learn a neural network as the bidder's supply function. The neural network $\pi_\theta$ is a mapping from market observations $o_t$ and market clearing price $\lambda_t$ to output power $p_t$ of the ESS. The relation between price $\lambda_t$ and power $p_t$ is the learned supply function.

We construct the reinforcement learning problem by defining a Markov Decision Process(MDP), which is characterized by the state space, action space, reward function, transition probability, and discount factor $(\mathcal{S}, \mathcal{A}, \mathcal{R}, \mathcal{T}, \gamma)$:

1. State $\mathcal{S}$: The state vector includes the energy clearing price $\lambda_t$ and the state observations $o_t = [T, H, SoC_t]$, which are the latest market information by bid submission time.

a. $\mathbf{T}$: $\sin(T/24h), \cos(T/24h)$, 2-dimensional sinusoidal encoded time to indicate time of day
b. $\mathbf{H}$: $(h_1,..,h_6,\overline{h_1},..,\overline{h_6})$, encoding of the market history, which consists of the angle and amplitude of the first three dimensions in the discrete Fourier Transform of the past 6 hours and past 4 days.
c. $SoC_t$: the State of Charge at $t$

2. Action $\mathcal{A}$: In the action space, we specifically define the price range for zero power output to facilitate energy withholding strategy. The ESS's power is determined by the output charging pair $(\lambda_t^c, p_t^c)$ and discharging pair $(\lambda_t^d, p_t^d)$:

$$p_t = \begin{cases} p_t^d & \text{, if } \lambda_t \geqslant \lambda_t^d \text{ (discharging)} \\ -p_t^c & \text{, else if } \lambda_t \leqslant \lambda_t^c \text{ (charging)} \\ 0 & \text{, else} \end{cases} \quad (5)$$

3. Reward $\mathcal{R}$: The net income reward (4) is used as the reward function.
4. Transition $\mathcal{T}$: The model is update with the ESS and the market dynamics.

The policy network is a multi-layer perceptron network that has two hidden layers of 256 nodes. All output values are mapped through Tanh activations.

### 3.3 Soft Constraint for Monotonicity

To ensure the monotonicity of the neural network supply function, we impose the soft constraint by proposing a new monotonicity loss.

The original monotonicity constraint is the nonnegativity of the partial derivative of the supply function, i.e. $\frac{\partial p}{\partial \lambda} \geq 0$. To impose a soft constraint, violated derivatives are penalized with the following loss function:

$$L_{mono} = (\frac{\partial p}{\partial \lambda})^2 \cdot (\frac{\partial p}{\partial \lambda} < 0) \quad (6)$$

where the first term quadratically penalizes the derivatives, and the second term limits the penalization for non-monotonic samples. To introduce this soft constraint to actor training, we add a monotonicity training process after each RL update. For each batch, the market state observation $o_t$ is sampled from the rollout history, and the energy price $\lambda_t$ is sampled from the price distribution. The monotonicity loss (6) is then calculated and backpropagated.

### 3.4 Output Activation for Discreteness

To satisfy the discreteness constraint, we propose an output activation layer. The output activation layer is defined by reference level $p_p^1, \cdots, p^{10}$. The original continuous power output $p$ is mapped to its nearest output reference $\bar{p}$:

$$\pi_{p_p^1 \cdots p^{10}}^d(p) = \arg\min_{\bar{p}=p_{\min}, p^1 \cdots p^{10}} |\bar{p} - p| \quad (7)$$

In practice, we derive the output reference level $p^i$ from the KMeans clustering centroids of a continuous supply function policy. A supply function in Section 3.2 is first learned, and its output actions are collected as a dataset for clustering.

With the modifications in Section 3.3 and 3.4, we can re-train the neural network for NNEB. With the two constraints satisfied, the neural network $\pi_{\theta,o_t}$ uniquely corresponds to a high-dimensional bid. However, due to length limitations, the method for extracting bids from $\pi_{\theta,o_t}$ will be omitted in this article.

## 4. PERFORMANCE EVALUATION

### 4.1 Experiment Settings

We conduct experiments on $\pm 1MW$ ESS units in the PJM market[3] with a default storage capacity of 8 MWH. We choose the PPO RL algorithm[8] for training and use 700 days of market history for training. The following 300 days are used for testing. The action references $p^i$s are derived by clustering under the same training condition. The training parameters can be found in *Table. 1*.

| Parameters | Value | Parameters | Value |
|---|---|---|---|
| Storage capacity | $8\ MWH$ | Number of envs | 256 |
| Power capacity | $1MW$ | $\gamma$ | 0.9999 |
| $\lambda^{dep}$ | 10 | $lr_\pi/lr_V$ | $3e^{-4}/1e^{-3}$ |
| $\eta_c/\eta_d$ | 0.95 | $\epsilon_{clip,PPO}$ | 0.2 |
| Reference levels | $-0.92/-0.83/-0.72/-0.61/-0.52/0/0.32/0.63/0.87/1$ | | |

*Table. 1 Training parameters*

### 4.2 Result Overview

We trained the actor for $5 \times 10^6$ steps on the training dataset. A demonstration of bidding is shown in *Fig. 2*. It includes the bidding result across 5 days. We can see that even though the market price highly fluctuates, the actor can manage its SoC to capture the peak prices. From the perspective of charging, it is able to feed the SoC to proper levels before peak times, and from the perspective of discharging, it is able to precisely capture the peak prices using high dimensional energy bids.

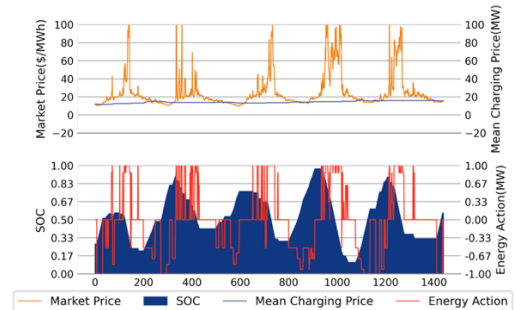

*Fig. 2 Bidding result of the NNEB*

In Section 3.3, a soft constraint is used to penalize non-monotonic network segments. With such a soft constraint, the trained actor is 99.23% monotonic on the testing dataset, which is practically feasible for generating market bids.

*4.3 Performance of NNEB*

We compare the performance of NNEB with other forms of bid parameterizations under the same training conditions. The action space and equivalent bids are described in *Table. 2*.

| Bid Type | Description |
|---|---|
| SELF | $p$ : quantity-only bids for self-scheduling |
| 2PAIR | $(p_c, \lambda_c), (p_d, \lambda_d)$: two price threshold bids and quantity bids for charging and discharging |
| NNEB | $(p_1, .., p_{10}, \lambda_1, .., \lambda_{10})$: full market bids |

*Table. 2 Formulation of different bid types*

*Fig. 3* shows the training curve of these methods. We can see that due to the high volatility in real-time market prices, SELF dispatch cannot capture the peak prices precisely and converges poorly. 2PAIR bidding can partially respond to market prices through price thresholds $\lambda_c, \lambda_d$ and achieves higher income. Finally, the NNEB actor is able to utilize the full bid space and achieves the highest bidding performance.

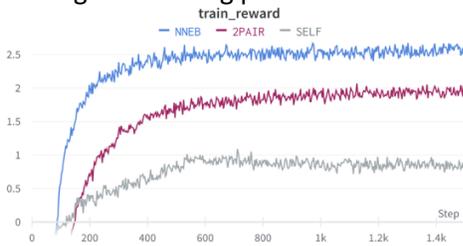

*Fig. 3 Training reward histories*

The testing results under different storage capabilities are also shown in *Table. 3*. We can see that NNEB generally achieves 18% higher profit than the 2PAIR bidding method. This shows that higher flexibility in bids enables better learning of strategies and higher income.

| Bid Type | 2hr | 4hr | 6hr | 8hr | 10hr | 12hr |
|---|---|---|---|---|---|---|
| Optimal | 2.275 | 2.497 | 2.588 | 2.627 | 2.646 | 2.659 |
| NNEB | 72.04% | 72.77% | 73.26% | 76.51% | 78.31% | 77.17% |
| 2PAIR | 45.27% | 54.51% | 55.22% | 56.34% | 59.79% | 59.20% |
| SELF | 23.70% | 32.35% | 32.66% | 33.91% | 34.58% | 35.17% |

*Table. 3 Testing performance of different bid types*

## 5. CONCLUSION

In this paper, a bid representation method called NNEB is proposed for representing high-dimensional market bids with monotonic and discrete neural networks. With such representation, the ESS bidding problem is factorized to supply function learning with monotonicity and discreteness constraints. A MDP is proposed to achieve efficient learning of a neural network supply function with RL algorithms. The monotonicity loss and the output activation layer are proposed to make the network monotonic and discrete. Finally, NNEBs can be learned and generated efficiently. The experiment results show that our proposed method enhances market bidding performance by 18% compared to the baseline, showing the importance of utilizing high-dimensional bids.

In the future, we will work on applying the bidding framework to various market types and participant types with more consideration of the agent interactions and market history.

## ACKNOWLEDGEMENT

This work has been supported by the National Natural Science Foundation of China in Grant 52107102